\documentclass[pra,aps,twocolumn,superscriptaddress,longbibliography,floatfix,notitlepage]{revtex4-2}
\usepackage[english]{babel}
\usepackage{lmodern}
\usepackage{amsmath,amsfonts,amssymb,amsthm,bm,times,dcolumn,dsfont}
\usepackage{microtype}
\usepackage{braket}
\usepackage{gensymb} 
\usepackage{physics}
\usepackage[colorlinks={true}, citecolor={blue}, filecolor={blue}, linkcolor={blue}, urlcolor={blue}]{hyperref}
\usepackage{graphicx,color}
\usepackage[caption=false]{subfig}
\begin{document}

\preprint{APS/123-QED}

\title{Macroscopic distant magnon-modes entanglement via a squeezed drive}%
\author{Kamran Ullah}
\email{kullah19@ku.edu.tr}
\affiliation{Department of Physics, Ko\c{c} University, 34450 Sar\i yer, Istanbul, T\"urkiye}
\author{M. Tahir Naseem} 
\email{mnaseem16@ku.edu.tr}
\affiliation{Faculty of Engineering Science, Ghulam Ishaq Khan Institute of Engineering Sciences and Technology, \\
Topi 23640, Khyber Pakhtunkhwa, Pakistan}
\author{\"Ozg\"ur E. M\"ustecapl\i o\u glu}	
\email{ omustecap@ku.edu.tr}
\affiliation{Department of Physics, Ko\c{c} University, 34450 Sar\i yer, Istanbul, T\"urkiye}
\affiliation{T\"{U}B\.{I}TAK Research Institute for Fundamental Sciences, 41470 Gebze, T\"{u}rkiye}

\date{\today}
\begin{abstract}
 The generation of robust entanglement in quantum system arrays is a crucial aspect of the realization of efficient quantum information processing. Recently, the field of quantum magnonics has garnered significant attention as a promising platform for advancing in this direction. In our proposed scheme, we utilize a one-dimensional array of coupled cavities, with each cavity housing a single yttrium iron garnet (YIG) sphere coupled to the cavity mode through magnetic dipole interaction. To induce entanglement between YIGs, we employ a local squeezed drive, which provides the necessary nonlinearity for entanglement generation. Our results demonstrate the successful generation of bipartite and tripartite entanglement between distant magnon modes, all achieved through a single quantum drive. Furthermore, the steady-state entanglement between magnon modes is robust against magnon dissipation rates and environment temperature. Our results may lead to applications of cavity-magnon arrays in quantum information processing and quantum communication systems.
\end{abstract}

\maketitle

\section{Introduction}

Quantum entanglement plays a pivotal role in various applications of quantum information processing \cite{Flamini_2019}, encompassing quantum cryptography~\cite{PhysRevLett.84.4729}, quantum teleportation \cite{PhysRevLett.123.070505}, and quantum metrology \cite{PhysRevLett.113.250801}. It is a crucial resource for enhancing the performance of quantum devices and technologies. However, preparing long-lived entangled states, especially at the macroscopic scale, becomes challenging because of the inevitable interactions between quantum systems and their environments. Consequently, the quest for generating macroscopic entangled states using various physical setups has gained significant attention. In this context, steady-state entanglement between atomic ensembles has been successfully demonstrated \cite{Julsgaard2001,Chou2005,McConnell2015}. Additionally, entangled states involving macroscopic systems have been reported in different setups, such as entanglement between a single cavity mode and a mechanical resonator in an optomechanical configuration \cite{science.1244563}. Recently, experimental achievements include entanglement between two macroscopic resonators coupled to a common cavity mode through optomechanical interaction \cite{Ockeloen-Korppi2018}.

Another central challenge in harnessing entanglement to improve various quantum tasks lies in the ability to create and distribute entanglement across large arrays of quantum systems \cite{Diehl2008,Verstraete2009,Weimer2010,Barreiro2011,PhysRevLett.106.020504,PhysRevLett.115.200502,PhysRevLett.117.040501}. An intriguing approach to address this challenge is reservoir engineering, where external control drives are used to engineer desirable dissipative dynamics, leading the quantum system to relax in the desired quantum state \cite{PhysRevLett.77.4728, PhysRevLett.86.4988, Barreiro2011, PhysRevLett.107.080503,Shankar2013,science.1261033,PhysRevLett.115.243601, Naseem2021, Naseem_2022}. However, most reservoir engineering schemes require the application of multiple external control fields on different elements of the quantum system array \cite{PhysRevLett.70.556,PhysRevA.88.063833}.

In recent years, hybrid quantum systems, combining different physical subsystems, have received significant attention \cite{TABUCHI2016729,Lachance-Quirion_2019,YUAN20221}. Among these systems, cavity magnon setups offer a unique platform to investigate light-matter interactions \cite{PhysRevLett.113.156401, PhysRevLett.114.227201}. In cavity quantum electrodynamics, the strong and ultra-strong coupling between magnons and photons has been increasingly studied \cite{PhysRevLett.113.156401, PhysRevB.93.144420, PhysRevLett.117.123605, PhysRevLett.117.133602, PhysRevLett.116.223601}, with experimental demonstrations of these regimes \cite{PhysRevLett.123.107702, PhysRevLett.123.107701,sciadv.abe8638}. This strong coupling opens possibilities for exploring novel phenomena, such as the magnon Kerr effect  \cite{PhysRevB.94.224410}, cavity-magnon polaritons \cite{Zhang2017}, magnon-induced transparency \cite{sciadv.1501286}, magnomechanically induced slow-light \cite{PhysRevA.102.033721}, bistability \cite{PhysRevLett.120.057202}, exceptional points \cite{Zhang2017, PhysRevB.99.054404}, magnon blockade~\cite{PhysRevA.101.042331}, and nonreciprocity \cite{PhysRevApplied.12.034001}. 

Recent proposals aim to realize genuine quantum effects in macroscopic magnon-cavity systems, including magnon blockade \cite{PhysRevB.100.134421}, magnon squeezed states \cite{PhysRevA.99.021801}, Schr\"odinger cat states \cite{PhysRevB.103.L100403, PhysRevLett.127.087203}, Bell states \cite{PhysRevLett.124.053602}, and nontrivial bipartite and multipartite entangled states \cite{PhysRevResearch.3.023126, PhysRevLett.121.203601}. Of particular interest is the generation of entanglement between distant magnon modes \cite{Luo:21}, which holds promise for quantum information processing applications. However, to the best of our knowledge, existing studies have primarily focused on entangling only two magnon modes, such as entangling two YIG spheres in a single cavity via Kerr nonlinearity \cite{PhysRevResearch.1.023021}, magnetostrictive interaction \cite{Li_2019}, or applying a vacuum-squeezed drive \cite{5.0015195, Zheng2023}. Further, stationary quantum entanglement between two massive magnetic spheres can be induced by subjecting each sphere to two-tone Floquet fields effectively generates parametric interaction between magnon modes~\cite{Xie_2023}. In the case of two cavities, each containing a single YIG, entanglement between the YIGs has been generated through optomechanical-like coupling \cite{Zhao2022}, vacuum squeezed drive \cite{Yu_2020}, or reservoir engineering \cite{Luo:21,PhysRevA.104.023711, PhysRevB.105.094422}. Recently, bipartite entanglement between magnon modes has been investigated in a one-dimensional array of cavities. The entanglement is generated by exploiting the coupling of the magnon modes with a central giant atom via virtual photons \cite{LIU2023106854}. 

\begin{figure*}[t]
\centering
\includegraphics[width=\linewidth]{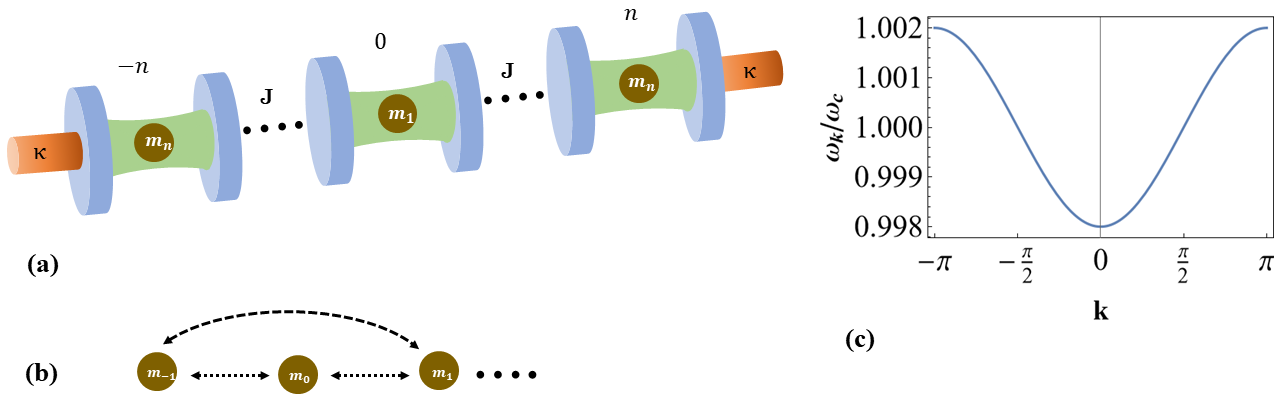}\label{fig:fig1a}
\caption{(a) Schematic description of magnon-magnon entanglement generation scheme, consisting of a one-dimensional $2N+1$ array of identical lossy cavities coupled via hopping interaction $J$. Each cavity contains a yttrium iron garnet (YIG) sphere with volume $V$. 
The central YIG is coupled to a quantum squeezed drive, enabling our entanglement generation scheme. (b) describes the effective model, which consists of only magnon modes obtained after removing the cavities via employing the Schrieffer-Wolf transformation. (c) shows the behavior of the dispersion relation of the cavity mode in momentum space.} \label{fig:fig1}
\end{figure*} 

Hybrid quantum magnonic systems hold great promise for efficient quantum networks \cite{PRXQuantum.2.040344}. Therefore, a crucial task is to establish and control entanglement between multiple magnon modes in arrays based on quantum magnonic systems. In this regard, we propose a scheme to generate entanglement between multiple yttrium iron garnet (YIG) spheres using a reservoir engineering approach with a single squeezed drive. Specifically, our setup consists of an array of cavities, each housing a macroscopic YIG coupled to the cavity through beam splitter-like interaction. However, entanglement cannot be generated with this interaction alone; thus, we introduce the necessary nonlinearity by driving one of the magnon modes with a squeezed drive. Previous studies have demonstrated that a single and two-mode squeezed drive can create entanglement between magnon modes \cite{Yu_2020,5.0015195, Zhao2022,Zheng2023}. In our work, we illustrate that subjecting one of the magnon modes to a squeezed drive is sufficient to generate steady-state bipartite and tripartite entanglement between distant magnon modes.  
We note that entanglement in a one-dimensional chain of coupled harmonic oscillators was proposed in Ref. \cite{PhysRevA.92.032319}. In that study, entanglement is generated only between pairs of modes symmetrically positioned relative to the central oscillator, which is coupled to the squeezed reservoir (see Fig. 2(a) in \cite{PhysRevA.92.032319}). In contrast, our scheme enables entanglement between all possible pairs in the chain, due to the all-to-all coupling between the magnon modes. Furthermore, tripartite entanglement among magnon modes has not been reported in previous studies \cite{PhysRevA.92.032319, Zheng2023}. Furthermore, we show that the entanglement between these distant magnon modes can be increased by more than one order of magnitude by applying two-photon drives to each cavity in the array.

In principle, we would like to highlight that the local squeezed drive on the central magnon mode ($m_{2}$) can be realized by coupling an auxiliary quantum system. Generation of the squeezed input field can be accomplished using a method similar to the proposal discussed in~\cite{PhysRevA.92.032319}. For example, the input squeezed drive can be engineered by coupling an auxiliary microwave cavity to the central magnon mode through beam-splitter-type interaction. The weak squeezed vacuum field, generated via a flux-driven Josephson parametric amplifier, acts as the driving force for the auxiliary cavity. Consequently, the auxiliary cavity serves as a quantum squeezed drive for the central magnon mode~\cite{PhysRevA.99.021801, PhysRevX.7.041011}.
Similarly, the squeezed drive can also be achieved by coupling the central magnon mode to a superconducting qubit~\cite{science.aaa3693}. More specifically, the approach proposed in ~\cite{PhysRevLett.108.043602} can be implemented by replacing the coplanar waveguide resonator with a YIG placed inside a microwave cavity. This scheme enables the coupling of a superconducting qubit with an adjustable energy gap to a magnon mode characterized by the frequency $\omega_2$. By modulating the energy gap of the qubit with a bichromatic field, the squeezed drive for the magnon mode can be engineered~\cite{PhysRevA.99.021801}. Alternatively, the input squeezed radiation can be generated by applying a two-tone microwave field drive to the central YIG, where the magnon mode is coupled to its mechanical vibrational mode through magnetostrictive interaction~\cite{Zhang:21,nwac247}. 

The paper is structured into the following sections: In Sec.~\ref{sec:model}, we introduce the model; and in Sec.~\ref{sec:effHam} derivation of effective Hamiltonian is presented. The results for bipartite and tripartite entanglement between the magnon modes are discussed in Secs.~\ref{subsec:bipartite} and \ref{subsec:tripartite}, respectively. In Sec. \ref{sec:conclusion}, we provide concluding remarks and summarize the key findings. In Appendix \ref{App:A}, we discuss bipartite and tripartite entanglement in the case of five cavities containing YIG spheres. Finally, we compare the generation of entanglement in cavities driven by squeezed drives with that in undriven cavities in Appendix \ref{App:B}.

\section{The Model}\label{sec:model}
We investigate a cavity-magnon system comprising a one-dimensional array of $2N+1$ cavities, as illustrated in Fig.~\ref{fig:fig1}. These cavities are interconnected through photon-hopping interactions. Inside each cavity, a single YIG sphere is present, coupled to the cavity mode via magnetic dipole interaction. In this study, we consider a single magnon mode, which represents quasi-particles with collective spin excitations, associated with each YIG in the cavities. The Hamiltonian governing the field for the cavity array is as follows:
\begin{align}
 H_a/\hbar =& \omega_c\sum_{j}\hat{a}^{\dagger}_j\hat{a}_j 
 -J\sum_{j}(\hat{a}_j\hat{a}_{j+1}^\dag + \hat{a}_j^\dag\hat{a}_{j+1}).
  \label{eq:hamilarray} 
\end{align}
The first term represents the free energy of the cavities, while the second term accounts for the exchange energy of the field. $\hat{a}_j$ and $\hat{a}_j^\dagger$ are the bosonic annihilation and creation operators of the $j$-th cavity and the corresponding resonance frequency $\omega_{c}$. Here, $j$ ($j = -N,.., 0, ..., N$) describes the index of the cavity in a one-dimensional array, with each cavity interconnected through a hopping coupling called $J$. We consider only a single photon process in each cavity. 

The ferromagnetic sample (YIG) holds within it scattering spin waves, with the assumption that only the spatially uniform Kittel mode \cite{PhysRev.73.155} exhibits a pronounced interaction with photons within the cavity.
The free Hamiltonian for these magnon modes is given by
\begin{equation}\label{eq:hamilmagnon}
     H_m/\hbar = \sum_n\omega_{n}\hat{m}^{\dag}_n\hat{m}_n,
\end{equation}
here $\hat{m}_n$ ($\hat{m}^{\dag}_n$) is the annihilation (creation) operator of the magnon mode, and $\omega_{n}$ is the associated frequency of the mode. In addition, the index $n$ in the summation is an integer number such that $n\in [-N, N]$. In general, some cavities can be empty while some can contain YIG spheres.
For example, in Sec. \ref{sec:results}, we assume that three of the cavities are occupied while the rest remain empty. In this scenario, Eq. (\ref{eq:hamilmagnon}) includes only three terms for magnon modes, however, $2N+1$ cavities are present. The magnon frequencies can be calculated based on the applied magnetic fields $H_n$, and given by $\omega_n = \gamma H_{n}$, where $\gamma/2\pi=28$ GHz/T represents the gyromagnetic ratio. The interaction between the magnon and cavity modes is given by 
\begin{equation}\label{eq:hamilint}
H_I/\hbar = \sum_{n}  g_n(\hat{a}_n\hat{m}^{\dag}_{n} + \hat{a}^{\dag}_n\hat{m}_{n}),
\end{equation}
 here, $g_n$ is the coupling strength between the magnon and associated cavity modes and is given by $g_n=\zeta\gamma/2\sqrt{5\hbar\omega_{n}\mu_0 N/V}$. The volume of the cavity is given by $V$,  $N$ represents the total number of spins in the YIG, $\mu_0$ is the permeability of free space, and $\zeta$ describes the spatial overlap between the magnon and photon modes. We note that the interaction term in Eq. (\ref{eq:hamilint}) is obtained after performing the Holstein-Primakoff transformation in which collective spin operators are written in the form of bosonic magnon operators $\hat{m}$ ($\hat{m}^\dagger$) \cite{PhysRevResearch.1.023021}. In addition, the counter-rotating terms $\hat{m}\hat{a}, \hat{m}^\dagger\hat{a}^\dagger$ are ignored assuming the validity of rotating wave approximation~\cite{5.0015195}.

\section{The Schrieffer–Wolff Approximation and Effective Hamiltonian}\label{sec:effHam}
To generate entanglement between two magnon modes, a nonlinear interaction
such as effective parametric-type nonlinear coupling $(\hat{m}_1^\dagger\hat{m}_2^\dagger+\hat{m}_1\hat{m}_2)$ between the modes is required. This interaction can be achieved by introducing  a strong Kerr nonlinearity \cite{PhysRevResearch.1.023021} or nonlinear magnomechanical interaction \cite{Li_2019}. It is typically easier to engineer an effective beamsplitter-type coupling $(\hat{m}_1^\dagger\hat{m}_2+\hat{m}_1\hat{m}_2^\dagger)$ between the magnon modes. For instance, in a system with two YIG spheres placed inside a microwave cavity, it is possible to generate the desired coupling $(\hat{m}_1^\dagger\hat{m}_2+\hat{m}_1\hat{m}_2^\dagger)$ by carefully selecting the system parameters that allow adiabatic elimination of the cavity mode~\cite{PhysRevApplied.13.014053}. Another approach involves an array of three cavities, where the central cavity contains a qubit, and each end cavity houses a YIG sample~\cite{PhysRevB.105.094422}. By employing the dispersive regime and using the Schrieffer-Wolff (Frohlich-Nakajima) approximation \cite{PhysRev.79.845, 00018735500101254, BRAVYI20112793}, the cavity modes can be eliminated, resulting in an effective beam splitter-type coupling between the magnon modes.  In our case, we adopt the latter method to create an array of YIGs with an effective coherent coupling $(\hat{m}_n^\dagger\hat{m}_{n+1}+\hat{m}_n\hat{m}_{n+1}^\dagger)$. Our strategy is to utilize these easier-to-engineer beam-splitter-type magnon-magnon interactions but introduce the required nonlinearity for generating entanglement between distant magnon modes through a squeezed thermal bath  \cite{PhysRevA.92.032319, Ma_2017, PhysRevResearch.2.023177, PhysRevLett.126.020402,Angeletti_2023}.

The Hamiltonian $H_a$ (Eq.~(\ref{eq:hamilarray})) associated with a one-dimensional array of coupled cavities represents the tight-binding bosonic model. To diagonalize it, we introduce new operators in the momentum space ($k$-space). The resulting diagonal Hamiltonian is given by
\begin{equation}
  H_{a}/\hbar = \sum_{k}\omega_k\hat{a}^{\dagger}_k\hat{a}_k,
\end{equation}
here, we have introduced
\begin{align}
    \hat{a}_k =& \frac{1}{\sqrt{2N+1}}\sum_{j}\hat{a}_j e^{ikj},\\
    \hat{a}^{\dagger}_{k} =& \frac{1}{\sqrt{2N+1}}\sum_{j}\hat{a}^{\dagger}_{j} e^{-ikj}.
    \label{eq:msoperator}
\end{align}
We have assumed the periodic boundary conditions such that $k:= k_m = 2 \pi m /(2 N +1)$ with $m \in [-N, N]$, and considering a large cavity array ($N \gg 1$) results in $k \in [-\pi, \pi]$. Moreover, $\omega_k=\omega_c-2J\text{cos}k$ is the dispersion relation of the cavity mode. The interaction Hamiltonian, as presented in Eq.~(\ref{eq:hamilint}), modifies to
\begin{equation}
H_I/\hbar =\frac{1}{\sqrt{2N+1}} \sum_{k,n}\left[g_n(\hat{a}_k\hat{m}^{\dag}_{n} e^{ik n} +\hat{a}_k^{\dag}\hat{m}_{n} e^{-ik n})\right].
\label{eq:int} 
\end{equation}
We note that $n$ is associated with the position of the YIG in the array (see Fig.~\ref{fig:fig1}). To derive an effective Hamiltonian involving only magnon modes, the elimination of cavity modes is necessary. This can be accomplished by invoking the Schrieffer-Wolf (Frohlich-Nakajima) transformation~\cite{PhysRev.79.845, 00018735500101254, PhysRev.97.869, PhysRev.149.491}. 
\begin{equation}
 [S, H_0]=-H_I.
\end{equation}
 Where $H_0$ is the free Hamiltonian consisting of cavity ($H_\text{diag}$), and magnon ($H_m$) modes. In addition, $H_I$ is the interaction Hamiltonian between the cavity and magnon modes, as given in~Eq.~(\ref{eq:int}).   
 $S$ is called a generator and it is anti-Hermitian in nature i.e., $S^\dagger =-S$, and it is given by
\begin{equation}
 S = \sum_{k, n^\prime}\left[\alpha_k^{n^\prime}\hat{a}_k\hat{m}_{n^\prime}^\dag - {\alpha_k^{n^\prime}}^\star \hat{a}^\dagger_{k}\hat{m}_{n^\prime}\right],
 \label{eq:generator1}
 \end{equation}
 here $\alpha_k^{n^\prime}=g_{n^\prime}/(\sqrt{2N+1}(\omega_{n^\prime}-\omega_k)) e^{-ik l_{n^\prime}}$.
The effective Hamiltonian of the system can be determined by unitary transformation $e^{S} H e^{-S}$. In the dispersive regime, such that $\omega_{n^\prime}-\omega_k\gg g_{n^\prime}/\sqrt{2N+1}$, we can ignore higher order terms in the expansion of the unitary transformation, i.e., $H_0 + 1/2[H_I, S]$, and keep terms up to the second order in $\alpha_k^{n^\prime}$. Consequently, the approximate effective Hamiltonian, comprising solely of magnon modes, can be explicitly expressed as follows \cite{PhysRevX.6.011032, PhysRevB.105.094422, PhysRevA.108.033717,LIU2023106854}:
\begin{align}
 H_\text{eff}/\hbar=& \sum_{n}\omega^{\prime}_n\hat{m}_n^\dag\hat{m}_{n} + \sum_{n, n^{\prime}}\mathcal{G}_{n n^{\prime}}(\hat{m}_n^\dag\hat{m}_{n^{\prime}} + \hat{m}_n\hat{m}^\dag_{n^{\prime}}).
 \label{eq:hamileff1}   
\end{align} 
 Where we have replaced discrete modes with a continuous distribution
 \begin{equation}
   \frac{1}{2N+1} \sum_k=\frac{1}{2\pi}\int_{-\pi}^{\pi} dk,  
 \end{equation}
in addition, the following integral identity is employed in the derivation of Eq.~(\ref{eq:geff})
\begin{equation}
\int_{-\pi}^{\pi} \frac{dk}{2\pi} \frac{e^{-ilk}}{U + V\text{cos}k}=(-1)^{|l|}\sqrt{\frac{1}{U^2-V^2}}e^{-|l|\text{arccosh}(U/V)}.
\end{equation}
 The effective frequency $\omega^{\prime}_{n}$ of the $n$-th magnon mode is given by 
\begin{align}
\omega^{\prime}_n &=\omega_{n} + \frac{g^2_n}{\sqrt{\Delta_{n}^2+ 4 J\Delta_{n}}}, \text{and}\\
\mathcal{G}_{n,n^\prime} &= \frac{g_n g_{n^\prime} (-1)^{\mid {nn^\prime} \mid}}{2} \bigg(\frac{e^{-|{n n^\prime}|\text{arccosh}{(1+\Delta_{n^\prime}}/2J)}}{\sqrt{\Delta_{n^\prime}^2+ 4 J\Delta_{n^\prime}}}
+\nonumber \\  &\qquad\qquad\qquad 
\frac{e^{-|{nn^\prime}|\text{arccosh}{(1+\Delta_{n}}/2J)}}{\sqrt{\Delta_{n}^2+ 4 J\Delta_{n}}}\bigg),\label{eq:geff}    
\end{align}
is spatially dependent on effective coupling strength between the magnon modes. Further, $\Delta_{n^\prime} = \omega_{n^\prime}-\delta_c$, $\Delta_n = \omega_n-\delta_c$, and $\delta_c =\omega_c -2J$ is the lower bound frequency of the cavity mode. In addition, $nn^\prime = \abs{n - {n^\prime}}$ is the distance between $n$-th and $n^\prime$-th YIG placed inside the cavity array. In our numerical simulations, we consider identical magnon modes such that $\omega_{n}=\omega$. 
The effective Hamiltonian in Eq.~(\ref{eq:hamileff1}) takes on the form of a beam splitter for magnon modes, which can be used for state transfer between distant magnon modes, as discussed in the previous studies \cite{PhysRevB.105.094422,LIU2023106854}. We note that, in our case, the interaction between magnon modes is mediated by virtual photons; and when contrasted with actual photons, virtual photons possess the advantages of being non-propagating and non-radiative. Consequently, energy exchange facilitated by virtual photons results in entirely coherent dynamics devoid of dissipation \cite{PhysRevB.105.094422}.

\section{Results}\label{sec:results}

To illustrate the working principles of our scheme, we initially focus on a simple scenario in which only three cavities are occupied, each containing one YIG. The remaining cavities in the array are not occupied. Note that the placement of YIGs is not restricted to neighboring cavities; they can be located in any of the three cavities within the array. In addition, we consider a squeezed thermal drive coupled to the central YIG.
It is worth noting that this entanglement generation scheme remains applicable for arrays with arbitrary cavities lengths~\cite{PhysRevA.92.032319}. 

We employ the quantum Langevin equations to describe the system's dynamics. By working in the interaction picture through the unitary evolution operator $\hat{U}(t)=\text{exp}[-it(\sum_{n}\delta^{\prime}_{n}\hat{m}^{\dagger}_{n}\hat{m}_{n})]$, the equations of motion take the following form (for convenience, we will now omit the hat symbol from operators)
\begin{align}\label{eq:eom}
\dot{m}_{-1}=&-\kappa_{-1} m_{-1}-i\mathcal{G}_{-1,0} m_{0}-i \mathcal{G}_{-1,1} m_1 + \sqrt{2\kappa_{-1}} m^\text{in}_{-1},\nonumber\\
\dot{m_{0}}=&-\kappa_{0} m_{0}-i\mathcal{G}_{-1,0} m_{-1}-i \mathcal{G}_{0,1} m_{1} + \sqrt{2\kappa_{0}} m^\text{in}_{0},\nonumber\\
\dot{m_{1}}=&-\kappa_{1} m_{1}-i\mathcal{G}_{-1,1} m_{-1}-i \mathcal{G}_{0,1} m_{0} + \sqrt{2\kappa_{1}} m^\text{in}_{1}.
\end{align}
Here, $\delta^{\prime}_{n} = \omega_0 - \omega^{\prime}_{n}$ represents the detuning of the $n$-th magnon mode with an effective frequency $\omega^{\prime}_{n}$ from the input squeezed drive of frequency $\omega_{0}$. Furthermore, $\kappa_n$ is the dissipation rate of the $n$-th magnon modes, and $m^{in}_n$ represents the corresponding input noise operator. The input noise operator $m^\text{in}_{0}$ accounts for driving the central magnon mode through a squeezed vacuum field. It is characterized by zero mean and following correlation functions~\cite{PhysRevLett.56.1917, PhysRevA.105.063516}
\begin{align}\label{eq:corr}
\langle m^\text{in}_{0}(t) m_{0}^{\text{in}\dag}(t^\prime)\rangle=&(\mathcal{N}+1)\delta(t-t^\prime),\nonumber\\
\langle m^{\text{in}\dag}_{0}(t) m_{0}^\text{in}(t^\prime)\rangle=&\mathcal{N}\delta(t-t^\prime),\nonumber\\
\langle m^\text{in}_{0}(t) m_{0}^\text{in}(t^\prime)\rangle=&\mathcal{M}\delta(t-t^\prime),\nonumber\\
\langle m^{\text{in} \dag}_{0}(t) m_{0}^{\text{in} \dag}(t^\prime)\rangle=&\mathcal{M}^{*}\delta(t-t^\prime).
\end{align}\label{eq:snoise}
Here $\mathcal{N}=\sinh^2{r}+\bar{n}_{0}(\sinh^2r+\cosh^2{r})$ and $\mathcal{M}$= $e^{i\theta}\sinh{r}\cosh{r}(1+2\bar{n}_0)$ with $\theta$ and $r$ being the phase and squeezing parameter of the input squeezed drive, respectively. In addition, $\mathcal{N}$ and $\mathcal{M}$ account for the number of excitations and correlations, respectively. We observe that, in the limit of weak system-bath coupling and $g_{n^\prime}/(\sqrt{2N+1}(\omega_{n^\prime}-\omega_k))\ll 1$, the effect of Schrieffer-Wolf transformation on the dissipative terms in Eq. (\ref{eq:corr}) can be ignored \cite{PhysRevB.82.094511,PhysRevLett.130.213604}. Among these parameters, $\mathcal{M}$ plays the most influential role in entanglement generation. The equilibrium mean thermal magnon number can be determined by $\bar{n}_{0}(\omega_{0})= [\exp(\hbar\omega_0/K_B T_0)-1]^{-1}$. 

The input noise operators for the other two magnon modes are characterized by the following correlation functions
\begin{align}
\langle m^\text{in}_{\alpha}(t) m_{\alpha}^{\text{in}\dag}(t^\prime)\rangle=&(\bar{n}_{\alpha}+1)\delta(t-t^\prime),\nonumber\\
\langle m^{\text{in}\dag}_{\alpha}(t) m_{\alpha}^\text{in}(t^\prime)\rangle=&\bar{n}_{\alpha}\delta(t-t^\prime),
\end{align}\label{eq:tnoise}
here $\alpha= -1, 1$, and $\bar{n}_{\alpha}= [\exp(\hbar\omega_\alpha/K_B T_{\alpha})-1]^{-1}$.

In our scheme, we showcase the possibility of generating entanglement between all potential bipartitions of the magnon modes by driving the central magnon mode with an input quantum-squeezed field. The quadratures for both the magnon modes and the input noise operators are defined as follows $x_n=(m_n + m_n^\dag)/\sqrt{2}$, $y_n=(m_n-m_n^\dag)/\sqrt{2}i$, and $x^{in}_n=(m^{in}_{n} + m^{in\dag}_{n})/\sqrt{2}$, $y_n=(m^{in}_{n}-m^{in\dag}_{n})/\sqrt{2}i$, respectively. In terms of quadrature fluctuations, the quantum Langevin equation~(\ref{eq:eom}) can be rewritten as 
\begin{equation}
\dot{F}(t)= F(t)A + N(t),\label{eq:matrix}
\end{equation}
where $F(t)$=[$x_{-1}(t)$, $y_{-1} (t)$, $x_0(t)$, $y_0(t)$, $x_1(t)$, $y_1 (t)]^T$, and $N(t)$=[$x^{in}_{-1}(t)$, $y^{in}_{-1}(t)$, $x^{in}_{0}(t)$, $y^{in}_{0}(t)$, $x^{in}_{1}(t)$, $y^{in}_{1}(t)]^T$  denote the quadrature fluctuation vectors of magnon and input noise operators, respectively. The drift matrix $A$ is given by
\begin{align}
A = 
\begin{pmatrix}
-\kappa_{-1} & 0  & 0  & \mathcal{G}_{-1,0}  &  0  &  \mathcal{G}_{-1,1} \\
0 & -\kappa_{-1} & -\mathcal{G}_{-1,0}  &  0  &  -\mathcal{G}_{-1,1}  &  0 \\
0  & \mathcal{G}_{-1,0} & -\kappa_0 & 0   & 0 & \mathcal{G}_{0,1} \\
-\mathcal{G}_{-1,0} &  0 & 0 & -\kappa_0 & -\mathcal{G}_{0,1} & 0\\
0  &  \mathcal{G}_{-1,1}  &  0  &  \mathcal{G}_{0,1}  & -\kappa_1  &  0\\
-\mathcal{G}_{-1,1}  & 0   &  -\mathcal{G}_{0,1}  &  0  &  0 & -\kappa_1
\end{pmatrix}.
\end{align}

As the effective Hamiltonian of the magnon modes, as given in Eq.~(\ref{eq:hamileff1}), is quadratic, and the input quantum noise is Gaussian, as a result, the state of the system also remains Gaussian. The reduced state, consisting of three magnon modes, forms a continuous variable three-mode Gaussian state. This state can be entirely characterized by a $6\times 6$ covariance matrix $V$, which is expressed as
 \begin{equation}
 V_{ij}=\frac{1}{2}\left\langle F_i(t) F_j (t) + F_j(t) F_i (t)\right\rangle \quad i,j=1, 2,\dots,6.\label{eq:covar}
\end{equation}
The steady-state solution can be obtained by solving the Lyapunov equation
\begin{equation}
{AV + V A^T=-D }\label{eq:lyapunov}  
\end{equation}
where $D$ is the diffusion matrix and can be derived from the noise correlation matrix; $\langle N_i(t) N_j (t^{\prime}) + N_j(t^{\prime}) N_i (t)\rangle/2=D_{ij}\delta(t-t^\prime)$. It can be written as a direct sum of $D = D_1\oplus D_2 \oplus D_3$, with $D_{\alpha} = \text{diag}[\kappa_{\alpha}(2\bar{n}_{\alpha}+1), \kappa_{\alpha}(2\bar{n}_{\alpha}+1)]$, and
\begin{figure}[t!]
    \centering
    \includegraphics[width=\linewidth]{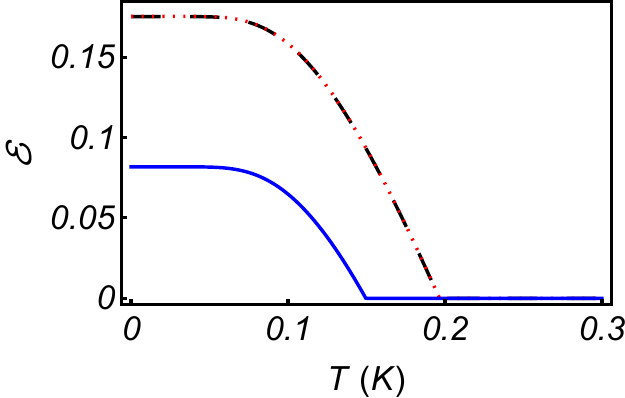}\label{fig:fig2}
  \caption{Magnon-magnon bipartite entanglement as a function of the environment temperature $T$. The black thicker-dashed line represents the logarithmic negativity $\mathcal{E}_{-1, 0}$, between $m_{-1}$ and $m_{0}$ modes, while the red dotted line indicates the logarithmic negativity $\mathcal{E}_{0, 1}$ between $m_{0}$ and $m_{1}$ modes. Further, the blue solid line shows the logarithmic negativity $\mathcal{E}_{-1, 1}$ between $m_{-1}$ and $m_{1}$ modes. Parameters: $\omega_c/2\pi=10$ GHz, $\omega_n/2\pi=3.331$ GHz, $g_j/2\pi= 10$ MHz, $J/2\pi=10$ MHz, $\kappa_{n}/2\pi$=$5$ MHz, and $r = 1$.}
\label{fig:fig2}
\end{figure}

 \begin{align}
D_2 = 
\begin{pmatrix}
\kappa_2\mathcal{U}_1  &  i \kappa_2 (\mathcal{M}^{*}-\mathcal{M})   \\
 i \kappa_2 (\mathcal{M}^{*}-\mathcal{M})  &  \kappa_2\mathcal{U}_2  
\end{pmatrix}.
\end{align}
where $\mathcal{U}_1=(2\mathcal{N}+ 1 + \mathcal{M} + \mathcal{M}^{*})$, and $\mathcal{U}_2=(2\mathcal{N}+ 1 - \mathcal{M} - \mathcal{M}^{*})$.

\subsection{Bipartite entanglement}\label{subsec:bipartite}

To investigate the entanglement between the magnon modes, we compute the logarithmic negativity $\mathcal{E}_N$. This quantity has previously been proposed as a measure of entanglement and helps establish the conditions under which the two modes are entangled \cite{PhysRevA.65.032314}. In the continuous variable case, the logarithmic negativity is defined as \cite{PhysRevA.70.022318}
\begin{equation}
 \mathcal{E} = \text{max}[0, -\ln 2 \mathcal{V_{-}}],
\end{equation}
where $\mathcal{V_{-}}=\sqrt{1/2(\Sigma-(\Sigma^2-4\det V^\prime)^{1/2}}$ with $\Sigma=\det A+\det B-2\det C$. $V^\prime$ is a $4\times4$ matrix obtained from steady-state covariance matrix $V$ by removing the two rows and associated columns related to the traced-out magnon mode. The reduced covariance matrix $V^\prime$ is given by
\begin{equation*}
V^\prime=
\begin{pmatrix}
A & C \\
C^T & B
\end{pmatrix}.
\end{equation*}

We compute the bipartite entanglement among all possible pairs of magnon modes by employing the logarithmic negativity $\mathcal{E}_N$. The analytical solution of equation~(\ref{eq:lyapunov}) is too cumbersome and we don't report it here. Instead, we extensively examine the logarithmic negativity across various system parameters. For numerical evaluations, we employ experimentally attainable parameters reported in recent studies~\cite{PhysRevLett.113.083603, PhysRevLett.116.240503,sciadv.1501286, PhysRevLett.129.123601, PhysRevLett.121.203601}. We consider the magnon density at low temperature with ground state spin $s=5/2$ of the $Fe^+_3$ ion in the YIG sphere. The total number of spins $N=\rho V$ with $\rho=4.22\times 10^{27}/m^3$ characterizing the density of each YIG and $V$ is the volume of each sphere with diameter 250 $\mu$-meter; this results in the total number of spins $N=3.5\times 10^{16}$ in each YIG. 
\begin{figure}[t]
 \centering
\includegraphics[width=\linewidth]{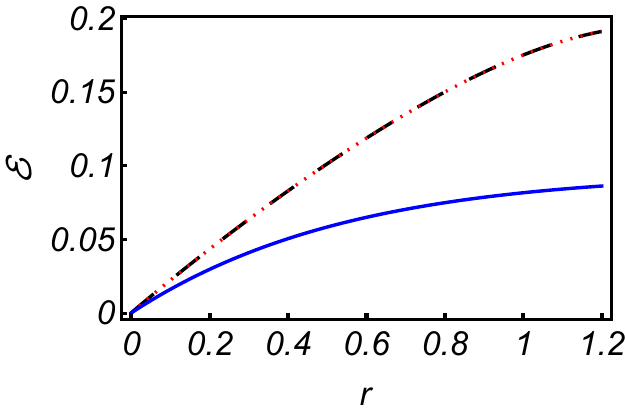}\label{fig:fig3}  
\caption{Magnon-magnon bipartite entanglement $\mathcal{E}$ as a function of squeezing parameter $r$. The black thicker-dashed line illustrates the logarithmic negativity $\mathcal{E}_{-1,0}$ characterizing the entanglement between $m_{-1}$ and $m_{0}$ magnon modes. The red dotted curve showcases the logarithmic negativity $\mathcal{E}_{0, 1}$ representing the entanglement between the $m_{0}$ and $m_{1}$ modes. Furthermore, the blue solid line shows the logarithmic negativity $\mathcal{E}_{1, 1}$ between $m_{-1}$ and $m_{1}$ considering the same parameters as those specified in Fig.~\ref{fig:fig2}.}\label{fig:fig3}
\end{figure}
 In Fig.~\ref{fig:fig2}, we present a plot illustrating bipartite entanglement, quantified using logarithmic negativity, among all possible pairs of magnon modes within an effective system comprising three magnon modes. These results are derived using a set of experimentally feasible parameters \cite{sciadv.1501286}: $\omega_n/2\pi=3.331$ GHz, $g_j/2\pi=10$ MHz, $J/2\pi=12$ MHz, $\omega_c/2\pi = 10$ GHz, $\kappa_{-1}/2\pi$=$\kappa_{1}/2\pi$= $5$ MHz, $r=1$, and the phase angle $\theta=0$. The entanglement is robust against temperature and survives up to about $T= 0.2$ K (black thicker-dashed line and red dotted line), and $T= 0.15$ K (blue solid line). The coupling $\mathcal{G}_{-1,1}$ between $\text{YIG}_{-1}$ and $\text{YIG}_{1}$ is relatively weak as compared to $\mathcal{G}_{-1,0}$ or $\mathcal{G}_{0,1}$ due to their farthest location and as a result, $\text{YIG}_{-1}$ and $\text{YIG}_{1}$ are relatively weakly entangled described by red line in Fig.~\ref{fig:fig2}. However, the more pronounced squeezing parameter $r$ would result in a more robust entanglement against the temperature of the environment. Here, we consider a reasonable value of the squeezing parameter $r=1$. 
 
 The logarithmic negativity $\mathcal{E}_{-1, 0}$ between $\text{YIG}_{-1}$ and $\text{YIG}_{0}$ and $\mathcal{E}_{0, 1}$ between $\text{YIG}_{0}$ and $\text{YIG}_{1}$ are the same due to the same couplings; $\mathcal{G}_{-1,0}$= $\mathcal{G}_{0,1}$ represented by black thicker-dashed and red dotted lines in Fig.~\ref{fig:fig2}.
The quantum squeezed drive coupled to the central magnon mode is responsible for entanglement generation between the magnon modes. The degree of entanglement decreases with a reduction in the squeezing parameter $r$ and dies in the absence of the squeezed drive.  We observe that the difference in the order of magnon decays at resonance frequencies can produce significant changes in the amount of steady-state entanglement between the two magnon modes. Each magnon-pair has a state-swap interaction and the squeezing can be transferred from a squeezed magnon state to another magnon state. As a result, each magnon pair is to be entangled due to swap-state type interaction via squeezing. This implies that squeezing can affect the bipartite entanglement generated in between a pair of the magnon modes. 

 To observe the impact of squeezing on entanglement generation, we plot the bipartite entanglement, quantified using logarithmic negativity ($\mathcal{E}$), as a function of the squeezing parameter ($r$). This can be observed in Fig.~\ref{fig:fig3} which indicates that in the absence of a squeezed thermal bath, the entanglement generation within our scheme is unattainable. Therefore, it is evident that squeezing plays a pivotal role in the generation of entanglement, with logarithmic negativity exhibiting a notable increase as the squeezing parameter ($r$) increases. The entanglement between the magnon modes $m_{-1}$-$m_{0}$ and $m_{0}$-$m_{1}$ is identical, owing to the equal coupling strength between these magnon pairs. However, the magnon modes $m_{-1}$-$m_{1}$ are relatively weakly coupled, primarily due to the greater distance between the associated YIGs. Consequently, the entanglement $\mathcal{E}_{-1, 1}$ between modes $m_{-1}$-$m_{1}$ is lower in comparison to the other two pairs. Entanglement generation in the case of five YIGs placed in a one-dimensional cavity array is discussed in Appendix \ref{App:A}.

\begin{figure}
\includegraphics[width=1\linewidth]{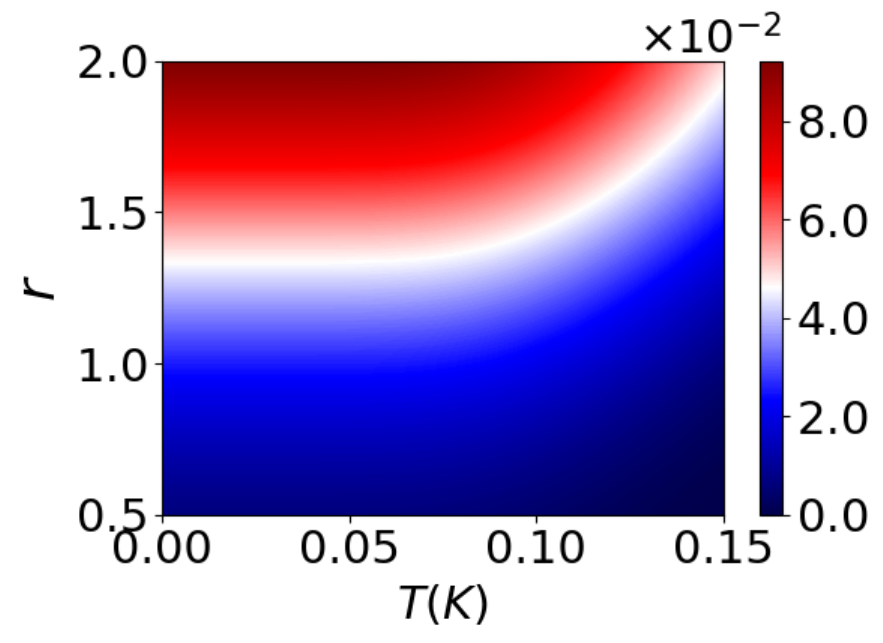}
\caption{Two-dimensional plot for tripartite entanglement, quantified by minimal residual contangle $\mathcal{R}$ as a function of bath temperature $T$, and squeezing parameter $r$. The rest of the system parameters are the same as given in Fig. \ref{fig:fig2}.}
\label{fig:fig4}
\end{figure}
\subsection{Tripartite entanglement}\label{subsec:tripartite}

We further investigate the possibility of the generation of the distant magnon-magnon tripartite entanglement in the array as shown in Fig.~\ref{fig:fig1}. To this end, we employ the minimal residual contangle given by \cite{Adesso_2006, Adesso_2007} 
\begin{equation}
    \mathcal{R}_{i|jk} = C_{i|jk} - C_{i|j} - C_{i|k}.
\end{equation}
Here, the contangle of subsystems $x$ and $y$ is denoted by $C_{x|y}$, and $y$ may represent more than one mode. $C_{x|y}$ is a proper entanglement monotone, and it is determined by the square of the logarithmic negativity between the respective modes, which is given by
\begin{equation}
    \mathcal{E}_{i|jk} = \text{max}[0, -\ln 2 \mathcal{V}_{i|jk}],
\end{equation}
where $\mathcal{V}_{i|jk}=\text{min}\abs{\text{eig}i\Omega_{3}\tilde{V}}$ is the smallest symplectic eigenvalues. $\Omega_{3}$ is the symplectic matrix $\Omega_{3} = \oplus^3_{j=1} i\sigma_{y}$, with $\sigma_{y}$ being the $y$-Pauli matrix and $\oplus$ symbol describes the direct sum of the $\sigma_{y}$ matrices. The $6\times 6$ covariance matrix $\Tilde{V}$ is obtained by inverting the momentum quadrature of one of the magnon modes. The transformed covariance matrix $\Tilde{V}$ is determined by $\Tilde{V} = P_{i|jk}VP_{i|jk}$ where $P_{1|23}=\text{diag}[1,-1,1,1,1,1]$, $P_{2|13}=\text{diag}[1,1,1,-1,1,1]$, and $P_{3|12}=\text{diag}[1,1,1,1,1,-1]$ are partial transposition diagonal matrices.
The steady state of the magnon modes can be fully characterized by the $6\times 6$ covariance matrix because of its Gaussian nature. The tripartite entanglement for Gaussian states can be determined by minimum residual contangle \cite{Adesso_2006, Adesso_2007} 
\begin{equation}
    \mathcal{R}_\text{min} = \text{min}[\mathcal{R}_{1|23}, \mathcal{R}_{2|13}, \mathcal{R}_{3|12}].
    \label{eq:contangle}
\end{equation}
This ensures that the tripartite entanglement remains unchanged regardless of how the modes are permuted.

The density plot of magnon-magnon tripartite entanglement determined via minimum residual contangle $\mathcal{R}_\text{min}$ is shown in Fig.~\ref{fig:fig4}. It is evident from the results that considerable tripartite entanglement can be generated between magnon modes when the central magnon mode is coupled to a squeezed drive. Fig.~\ref{fig:fig4} shows that the degree of entanglement is significantly enhanced with the increase in the squeezing parameter $r$.
\allowdisplaybreaks

\begin{figure}[h!]
    \centering
    \begin{minipage}[b]{0.95\linewidth}
        \centering
        \includegraphics[width=\linewidth]{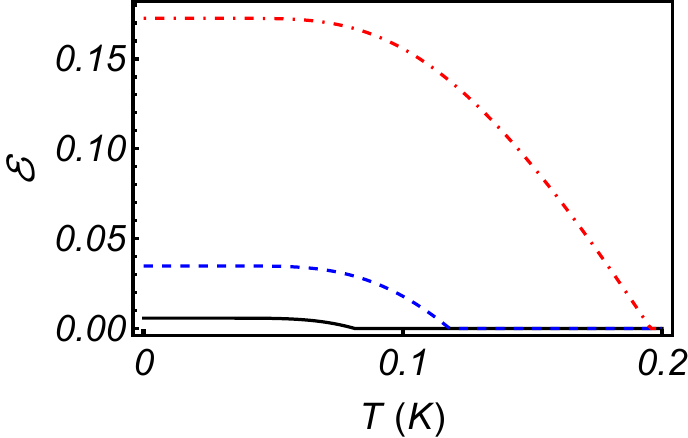}
    \end{minipage}
    \hspace{0.05\linewidth}
    \begin{minipage}[b]{0.95\linewidth}
        \centering
        \includegraphics[width=\linewidth]{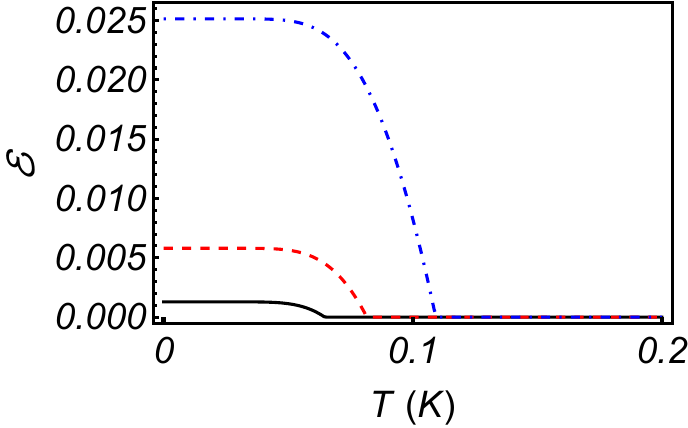}
    \end{minipage}
    \caption{The panels show bipartite entanglement between different pairs for a sample of five YIGs. (a) The black solid, blue dashed, and red dotdashed lines indicate the entanglement between $m_{-2}$ and $m_1$ modes, $m_{-1}$ and $m_1$, and $m_0$ and $m_1$, respectively. (b) Describes the bipartite entanglement between $m_{-2}$ and $m_2$ (solid black curve), $m_{-1}$ and $m_2$ (red dashed curve), $m_0$ and $m_2$ blue dotdashed Curve. The parameters remain the same as used in Fig.~\ref{fig:fig3}.}
    \label{fig:fig5}
\end{figure}

\section{Conclusion}\label{sec:conclusion}
In summary, we have shown that steady-state bipartite and tripartite entanglement can be generated between distant magnon modes when only one of these magnon modes is coupled to a quantum-squeezed drive. In particular, we considered an array of $N$ cavities, each of which houses a single YIG sphere, and the central YIG is coupled to a single-mode squeezed vacuum bath. We have demonstrated that steady-state bipartite and tripartite entanglement can be generated among magnon modes hosted by YIGs when some of the cavities, up to five, are occupied. In contrast to previous proposals for the generation of magnon-magnon entanglement \cite{Li_2019, 5.0015195, Yu_2020, Luo:21, PhysRevResearch.1.023021, Zhao2022, PhysRevA.104.023711, PhysRevB.105.094422, Zheng2023}, bipartite and tripartite entanglement are possible in our scheme. Further, in principle, our scheme can be extended to an arbitrary number of YIGs within the cavity array \cite{PhysRevA.92.032319, PhysRevLett.126.020402}. However, achieving entanglement across the entire array may necessitate negligible or extremely small magnon decay rates. To address this challenge, we propose the use of an external squeezed drive on each cavity in the array, enabling strong long-range interactions between YIGs positioned at greater distances within the array \cite{PhysRevA.108.033717}. The results presented in Appendix \ref{App:B} confirm that cavity modes driven by two-photon drives exhibit significantly enhanced entanglement compared to undriven cavities. In our scheme, the entanglement generation does not require nonlinearity, instead, entanglement is generated via a squeezed drive and its strength depends on the squeezing parameter $r$. The generated entanglement is robust against the environment temperature provided the magnon dissipation rates are sufficiently low. It may be interesting to look for multipartite entangled states of magnons, which is left for future works. Our work may be useful in designing quantum networks based on cavity-magnon systems \cite{PRXQuantum.2.040344}.

\appendix

\section{Bipartite entanglement with five YIGs}\label{App:A}

Here, we extend our results for five YIGs placed inside the cavities array, and the squeezed drive is coupled to the central YIG. The Langevin equations of motion in this case are given by
\begin{align}
\dot{m}_{-2}=&-\kappa_{-2} m_{-2} -i\sum_{n^\prime=-1}^{2}\mathcal{G}_{{-2}n^\prime}m_{n^\prime} + \sqrt{2\kappa_{-2}} m_{-2_{in}},\nonumber\\
\dot{m}_{-1}=&-\kappa_{-1} m_{-1}-i\mathcal{G}_{-2,-1} m_{-2} -i\sum_{n^\prime=0}^{2}\mathcal{G}_{-1n^\prime}m_{n^\prime}\nonumber\\  &+  \sqrt{2\kappa_{-1}}m_{-1_{in}}, \nonumber\\
\dot{m_0}=&-\kappa_0  m_0-i\sum_{n=-2}^{-1}\mathcal{G}_{n 0} m_{0}-i\sum_{n^\prime=1}^{2}\mathcal{G}_{0 n^\prime}m_{n^\prime} \nonumber\\  &+ \sqrt{2\kappa_0} m_{0_{in}},\nonumber\\
\dot{m_1}=&-\kappa_1 m_1-i\sum_{n=-2}^{0}\mathcal{G}_{n4} m_n- i\mathcal{G}_{12} m_2 + \sqrt{2\kappa_1 } m_{1_{in}},\nonumber\\
\dot{m_2}=&-\kappa_2 m_2-i\sum_{n=-2}^{1}\mathcal{G}_{n2} m_2 +\sqrt{2\kappa_2} m_{2_{in}}.
\label{eq:eom1}
\end{align}
Again we write the Langevin equations in matrix form as equation~(\ref{eq:matrix}). The A matrix is now a $10\times 10$ matrix which can be described as 
 \begin{widetext}
\begin{align}
A = 
\begin{pmatrix}
-\kappa_{-2} & 0 & 0 & \mathcal{G}_{-2,-1} &  0 & \mathcal{G}_{-2,0} & 0 & \mathcal{G}_{-2,1} & 0 & \mathcal{G}_{-2,2} \\
0 & -\kappa_{-2} & -\mathcal{G}_{-2,-1} &  0 & -\mathcal{G}_{-2,0} & 0 & -\mathcal{G}_{-2,1} & 0 & -\mathcal{G}_{-2,2} & 0 \\
0 & \mathcal{G}_{-2,-1} & -\kappa_{-1} & 0  & 0 & \mathcal{G}_{-1,0} & 0 & \mathcal{G}_{-1,1} & 0 & \mathcal{G}_{-1,2} \\
-\mathcal{G}_{-2,-1} &  0 & 0 & -\kappa_{-1} & -\mathcal{G}_{-1,0} & 0 & -\mathcal{G}_{-1,1} & 0 & -\mathcal{G}_{-1,2} & 0\\
0  &  \mathcal{G}_{-2,0}  &  0  &  \mathcal{G}_{-1,0}  & -\kappa_0 & 0 & 0 & \mathcal{G}_{0,1} & 0  & \mathcal{G}_{0,2} \\
-\mathcal{G}_{-2,0}  &  0   &  -\mathcal{G}_{-1,0}  &  0  &  0 & -\kappa_0 & -\mathcal{G}_{0,1} &  0 & -\mathcal{G}_{0,2} & 0 \\
0 & \mathcal{G}_{-2.1} & 0 & \mathcal{G}_{-1,1} & 0 & \mathcal{G}_{0,1} & -\kappa_1 & 0 & 0 & \mathcal{G}_{1,2} \\
-\mathcal{G}_{-2,1} & 0 & -\mathcal{G}_{-1,1} & 0 & -\mathcal{G}_{0,1} & 0 & 0 & -\kappa_1 & -\mathcal{G}_{1,2} & 0\\
0 & \mathcal{G}_{-2,2} & 0 & \mathcal{G}_{-1,2} & 0 & \mathcal{G}_{0,2} & 0 & \mathcal{G}_{1,2} &  -\kappa_2 & 0\\
-\mathcal{G}_{-2,2} & 0 & -\mathcal{G}_{-1,2} & 0 & -\mathcal{G}_{0,2} & 0 & -\mathcal{G}_{1,2} & 0 & 0  &  -\kappa_2
\end{pmatrix}.\label{eq:matrix10}
\end{align}
\end{widetext}
With $F(t)$=$[x_i(t)$, $y_i (t)]^T$ and $N(t)$=$[x_{i_{in}}(t)$, $y_{i_{in}}(t)]^T$ \quad i=1,2,3,\dots, 10.
We can write equation~(\ref{eq:matrix10}) for $N$ magnon modes in general $2N\times 2N$ matrix form in the interaction picture can be expressed as

\begin{figure}
\includegraphics[width=1\linewidth]{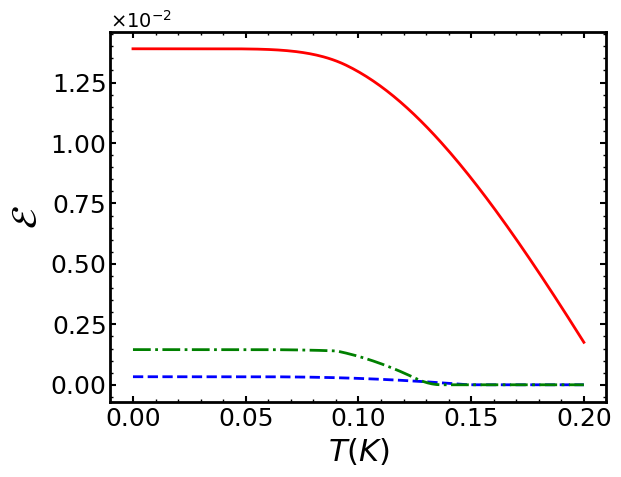}
\caption{Tripartite entanglement $\mathcal{E}$ between magnon modes in a cavity array containing five YIGs is shown as a function of temperature $T$. The solid, dot-dashed, and dashed lines represent the entanglement between the magnon modes $m_{-1}, m_0, m_1$; $m_{-1}, m_0, m_2$; and $m_{-2}, m_0, m_2$, respectively. We consider a squeezing parameter of $r = 1.4$, while the remaining system parameters are the same as those in Fig. \ref{fig:fig2}.}
\label{fig:fig6}
\end{figure}

\begin{equation*}
Q_{n n^\prime}=
\begin{pmatrix}
K_n & G_{n n^\prime}    \\
G_{n n^\prime} & K_n  
\end{pmatrix}
\end{equation*}
where, $K_n = \begin{pmatrix}
-\kappa_n & 0  \\
0 & -\kappa_n
\end{pmatrix}$,
 and $G_{nn^\prime}= \begin{pmatrix}
0 & \mathcal{G}_{n n^\prime}   \\
-\mathcal{G}_{n n^\prime} & 0
\end{pmatrix}$.
are  $2\times 2$ matrices with $n=1$ \dots $N-1$, and $n^\prime=2$,\dots $N$. The bipartite and tripartite entanglement in the case of five YIGs is shown in Figs. \ref{fig:fig5} and \ref{fig:fig6}, respectively. Our results can be extended to $N$ YIGs placed in the cavity array with a squeezed drive coupled to the central YIG. 
 
\section{Entanglement generation in case of driven cavities}\label{App:B}
The entanglement between magnon modes in distant cavities weakens due to the small coupling strength between these modes, as shown in Figs. \ref{fig:fig5} and \ref{fig:fig6}. This reduction in coupling strength, however, can be mitigated by driving each cavity mode with a squeezed drive \cite{PhysRevA.108.033717}. In this case, an additional term representing the squeezed drives on the cavity modes needs to be added to Eq. (\ref{eq:hamilarray}). The complete Hamiltonian of the system is then given by
\begin{equation}
\begin{aligned}
H = &\; \delta_a \sum_n a_n^\dagger a_n + \sum_n\omega_{n}\hat{m}^{\dag}_n\hat{m}_n - \left[ J \sum_n a_n^\dagger a_{n+1} \right. \\
& \left. - \zeta \sum_n a_n^{\dagger 2} e^{-i \phi} - \sum_{n}  g_n\hat{a}_n\hat{m}^{\dag}_{n}  + \text{H.c.} \right],
\end{aligned}
\end{equation}
Here, $\zeta$ and $\phi$ represent the driving amplitude and the associated phase of the squeezed drives, respectively. Additionally, $\delta_a = \omega_c - \omega_s/2$ denotes the detuning of each cavity mode with respect to the squeezed drive frequency $\omega_s$. To derive the effective Hamiltonian for the magnon modes, it is convenient first to perform Bogoliubov squeezing transformations on the cavity modes, which are given by
\begin{equation}
    a_n = \alpha_n \cosh(r) - \alpha_n^\dagger e^{-i \phi} \sinh(r),
\end{equation}
Here, the squeezing parameter is given by $r = 1/4 \ln[\left(\delta_a + 2\zeta)/(\delta_a - 2\zeta \right)]$. In this squeezed frame, the approximate Hamiltonian, obtained by dropping the non-resonant terms, is given by \cite{PhysRevA.108.033717}
\begin{equation}
\begin{aligned}
H = &\; \delta_s \sum_n \alpha_n^\dagger \alpha_n + \sum_n\omega_{n}\hat{m}^{\dag}_n\hat{m}_n - \left[ J_s \sum_n \alpha_n^\dagger \alpha_{n+1} \right. \\
& \left. - \sum_{n}  G_n\alpha_n\hat{m}^{\dag}_{n}  + \text{H.c.} \right],
\end{aligned}
\label{eq:bigloHamil}
\end{equation}
here, $\delta_s = \delta_a/\cosh(2r)$, $J_s = J \cosh(2r)$, and $G_n = g_n \cosh(r)$ are the modified detuning and coupling strengths, respectively. Following the procedure presented in Sec. \ref{sec:effHam}, the effective Hamiltonian of the magnon modes can be expressed in a form similar to that given in Eq. (\ref{eq:hamileff1}), since the Hamiltonian provided in Eq. (\ref{eq:bigloHamil}) is similar with what we considered for our model in Sec. \ref{sec:model}. Thus, we can utilize Eq. (\ref{eq:hamileff1}) for entanglement generation in the case of driven cavities, with the difference that the coupling strengths are replaced from $g_n$ and $J$ to $G_n$ and $J_s$, respectively. More specifically, the effective frequencies and coupling terms in Eq. (\ref{eq:hamileff1}) are now given by
\begin{align}
\omega^{\prime}_n &=\omega_{n} + \frac{G^2_n}{\sqrt{\Delta_{n}^2+ 4 J\Delta_{n}}}, \text{and}\\
\mathcal{G}_{n,n^\prime} &= \frac{G_n G_{n^\prime} (-1)^{\mid {nn^\prime} \mid}}{2} \bigg(\frac{e^{-|{n n^\prime}|\text{arccosh}{(1+\Delta_{n^\prime}}/2J_s)}}{\sqrt{\Delta_{n^\prime}^2+ 4 J\Delta_{n^\prime}}}
+\nonumber \\  &\qquad\qquad\qquad 
\frac{e^{-|{nn^\prime}|\text{arccosh}{(1+\Delta_{n}}/2J_s)}}{\sqrt{\Delta_{n}^2+ 4 J\Delta_{n}}}\bigg),\label{eq:geff}    
\end{align}
here, $\Delta_{n^\prime} = \omega_{n^\prime}-\delta_c$, $\Delta_n = \omega_n-\delta_c$, and $\delta_c =\delta_s -2J_s$ is the lower bound frequency of the cavity mode.
\begin{figure}
\includegraphics[width=1\linewidth]{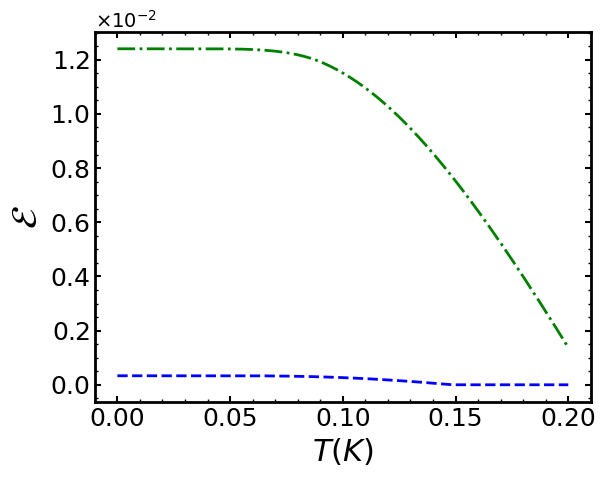}
\caption{The tripartite entanglement $\mathcal{E}$ between magnon modes in a cavity array containing five YIGs is presented as a function of temperature $T$. The dot-dashed and dashed lines represent the entanglement between the magnon modes $m_{-2}, m_0, m_2$ for cavities driven by squeezed drives and in the absence of these drives, respectively. We consider a squeezing parameter of $r = 1.4$, while the remaining system parameters are the same as those shown in Fig. \ref{fig:fig2}.}
\label{fig:fig7}
\end{figure}

We compare the tripartite entanglement based on the minimum residual contangle $\mathcal{R}_\text{min}$, as given in Eq. (\ref{eq:contangle}), for the case of a five-YIG array, both for non-driven (Sec. \ref{sec:model}) and driven cavities, in Fig. \ref{fig:fig7}. It is evident from the figure that the entanglement between comparatively distant modes is significantly larger—by more than one order of magnitude—in the case of cavities driven by two-photon drives compared to non-driven cavities. Therefore, incorporating squeezed drives, which enhance the effective coupling between magnon modes, can significantly improve entanglement generation. Our results may facilitate the generation of entanglement between distant magnon modes, which has not been reported in previous works \cite{Zheng2023}.

\bibliography{MagnonEntang}
\end{document}